 \documentstyle[aps,epsfig,float,amstex]{revtex}

\begin{document}
\title{Pairing in the Hubbard model: the Cu$_{5}$O$_{4}$ Cluster versus the Cu-O plane}
\author{Michele Cini, Adalberto Balzarotti, and Gianluca Stefanucci}
\address{Istituto Nazionale di Fisica della Materia, Dipartimento di Fisica,\\
Universita' di Roma Tor Vergata, Via della Ricerca Scientifica, 1-00133\\
Roma, Italy}
\maketitle
\begin{abstract}
 We study the Cu$_{5}$O$_{4}$ cluster by exact diagonalization of a 
 three-band Hubbard model and show that bound electron 
or hole pairs are obtained at appropriate fillings, and produce 
superconducting flux quantisation.
 The results extend
 earlier cluster studies and illustrate a canonical 
transformation approach to pairing that we have developed recently for 
the full plane. The quasiparticles that in the many-body problem 
behave like Cooper pairs are W=0 pairs, that is,  two-hole 
eigenstates of the Hubbard Hamiltonian with vanishing on-site 
repulsion.  The cluster allows W=0 pairs of $d$ symmetry, due to a spin
fluctuation, and $s$
symmetry, due to a charge fluctuation.  Flux 
quantisation is shown to be a manifestation of symmetry properties 
that hold for clusters of arbitrary size.
\end{abstract}


\noindent

\narrowtext

\twocolumn

\section{INTRODUCTION}

In many high-$T_{c}$ cuprates, one has superconductivity at 
concentrations about 1.2 holes/Cu atom, that is, somewhat 
above the antiferromagnetic region at half filling. In some cases, one 
finds 
superconductivity below half filling. Usually, one considers half filling as the vacuum and 
speaks of electron superconductivity in such cases. Electron pairing is actually 
realized \cite{uchida}, e.g., in the $T^{\prime }$
structure of (Nd,Ce)$_{2}$CuO$_{4}$. The $T^{\prime }$ structure of this compound is different from the
$T$ structure of La$_{2}$CuO$_{4}$, but is still characterized by CuO$_{2}$ planes \cite
{tp}.  However, increasing experimental evidence obtained in several cuprate superconductors
suggests that the pairs exist above the critical temperature
either in the form of superconducting fluctuations or preformed pairs. The latter
aspect is apparent in the underdoped (normal) region in which a clear
pseudogap essentially of the same magnitude as the superconducting gap is
measured\cite{b}. The pairing state of these materials has 
$d-wave$ symmetry, probably mixed with  $s-wave$\cite{a}. All these signatures put strict constraints to any
microscopic model of the cuprates. Any theory of the paired state must
predict $d$ and $s$ symmetries, and the pairing mechanism must be 
robust. It must survive well into the normal
state, and operate in a wide range of concentrations far from optimum doping.

In BCS theory, the first-order repulsion between like charges is overcome by
the second order interaction with phonons. In high-$T_{c}$ superconductors
the electron-phonon interaction is strong and phonons must be expected to
contribute in an important way to the pairing interaction, although their
task looks harder because the repulsion integral U is large (several eV).
However, the straightforward idea that the high-$T_{c}$ phenomena are just
a rescaled version of BCS theory is not granted. The role of phonons may be
important, but is different, and some other ingredient is essential.

Rather than proposing a new attractive interaction, we wish to point out that the 
popular repulsive Three-Band Hubbard model already leads to pairing. The model is: 
\begin{equation}
 H=H_{0}+W  \label{h}
 \end{equation}
 where the independent hole hamiltonian reads, in the site representation
 
 \begin{equation}
 H_{0}={\sum_{Cu}}\varepsilon _{d}n_{d}+{\sum_{O}}\varepsilon _{p}n_{p}+{\
 t\sum_{n.n.}}\left[ c_{p}^{\dagger}c_{d}+h.c.\right]  \label{h0}
 \end{equation}
 where n.n. stands for nearest neighbors\cite{gauge}. The on-site repulsion Hamiltonian
 will be denoted by 
 \begin{equation}
 W={\sum_{i}}U_{i}n_{i+}n_{i-},  \label{w}
 \end{equation}
 where $U_{i}=U_{d}$ for a Cu site, $U_{i}=U_{p}$ for an Oxygen.
 As in previous work\cite{cb1}, we use standard parameter values, $i.e.$, U$_{d}$=5.3,U$_{p}$= $6$,
t=1.3, $\varepsilon _{d}$=0, $\varepsilon _{p}$=3.5.
 The hole parameters $U_{d}=5.3$ eV, $U_{p}=6.$ eV differ somewhat from
other literature estimates\cite{mcm}, and must depend on the compound and
doping. For La$_{2}$CuO$_{4}$, $U_{p}$=4 eV and $U_{d}$=10.5 eV have been
recommended \cite{hyb}. None of our results depends qualitatively on
the precise value of the model parameters, since ours is basically a
symmetry argument. The electronic properties of this model are under intense 
investigation by several approximations based on perturbation theory 
and the Bethe-Salpeter equation. The  FLEX approximation is a
generalized RPA\cite{bsw} and leads to pairing and superconductivity in 
the three-band Hubbard model\cite{eb}. The excitation spectra of the 
2D Hubbard model have also been studied by a related self-consistent and 
conserving T-matrix approximation  by Dahm and  Tewordt\cite{dt}; 
we  mention incidentally that recently diagrammatic methods have been 
successfully applied to the photoelectron spectra of the Cuprates in other contexts 
too, like the spin-fermion model\cite{sps}.  A perturbative expansion around the strong coupling limit,
in powers of the kinetic energy, requires a nonstandard cumulant expansion, but 
is feasible, as shown quite recently by Citro and 
Marinaro\cite{citro1} for the $p-d$ model which is the present model 
with $U_{p}=0$. In this way, they have shown that normal state 
properties like the specific heat as a function of doping can be well 
understood\cite{citro2}; they also derived the effective pairing 
interaction in the same approximation\cite{citro3} and studied the doping 
dependence of the superconducting transition temperature \cite{t-j}.

Our starting point is the observation that, due to the planar C$_{4v}$
symmetry, there is actually no
repulsion barrier to overcome. In a series of papers\cite{cb1,cb2,cbpr} we have
introduced the two-hole singlet eigenstates of the Hamiltonian with zero Coulomb on-site
repulsion (the so called W=0 pairs). They arise in  the full plane, 
and also in the clusters that possess the same full C$_{4v}$ symmetry around a central Cu 
as the full plane.  In the full plane, this situation is always 
realised, because W=0 pairs can always be obtained from holes at the 
 Fermi level; in clusters, on the other hand, the hole number 
 (relative to the true hole vacuum) must be such that  two  
 holes partially fill a degenerate state. In the many-body problem, two holes at the Fermi level 
in a W=0 pair state do not interact directly; however the pair is dressed by the interaction with 
the background particles. By exact diagonalisation of cluster 
 Hamiltonians with up to 21 atoms and 4 holes, we demonstrated\cite{cbpr} that 
 the dressed W=0 pair is a bound Cooper pair, and quantizes the 
 magnetic flux like superconductors do. Any strong distortion 
 of the cluster symmetry breaks the pairing and restores the normal 
 repulsion\cite{cb2}. We also considered first-neighbor O-O hopping and off-site 
 interactions\cite{cb2}. Remarkably, the off-site repulsive interactions, when
included, tend to enhance the effect somewhat, so we devote the
present study to the on-site interaction effects. The 
binding energy of the pairs in these clusters is of the order of tens of meV, which is 
not comparable to any of the $U$ and $t$ input parameters. The reason 
is that the interaction, which vanishes identically for the {\em 
bare} W=0 pairs, remains {\em dynamically small} for the dressed 
quasiparticles. Indeed, by a  diagrammatic analysis we demonstrated 
that low-order perturbation theory is a good approximation to the exact 
diagonalisation results and allows to understand that the attraction 
in the d channel is due to virtual spin-flip
excitations. This suggests that a {\em weak coupling} theory may be 
useful to study the pairing force, despite the fact that $U$ is not
small compared to $t$. It is an obvious limitation of the cluster 
approach that W=0 pairs are possible at discrete values of the hole 
concentration. Our previous cluster calculations suggest that the mechanism operates in a
much broader range of hole concentrations than is realised in actual
Cuprates, from very highly overdoped (as in $CuO_{4}$) to very low
(as in $Cu_{5}O_{16}$). The diagrammatic analysis further demonstrates 
that the effective interaction is the result of a partial cancellation 
of positive and negative contributions, so it is not necessarily 
attractive in all cases; the general signature of W=0 pairs is that 
the absolute value of the interaction is much smaller than in the 
other cases.

The mechanism we are considering is only a part of the story, but it seems to be
a most peculiar part, being related to nothing but the C$_{4v}$ symmetry. 
For similar reasons here we wish to make abstraction from phonon effects to
see how far the idealised description can account for reality by itself. We
believe that a mechanism which predictably gets attraction out of repulsion
is by itself of theoretical interest.

Next, we have generalised the theory of pairing to the full plane\cite{csb}. 
In short, one finds W=0 pairs at the Fermi level for any
concentration and this leads to a Cooper-like instability of the Fermi 
liquid. Pairing prevails for a range of 
concentrations above half filling,in agreement with the results \cite{zanchi}
of the Renormalization Group technique.  
We have shown that the full configuration interaction calculation can 
be performed recursively. At each step, one decouples a class of 
virtual excitations while renormalising the matrix elements of $H_{0}$ 
and $W$. At the end, one obtains an exact, analytical canonical transformation 
producing an effective Hamiltonian for the dressed pair. 
In order to get actual numbers, however, we 
had to neglect the renormalisations in the final formula; 
this  approximation is fully justified at weak 
coupling. 

In the present paper we extend the analysis of Ref.\cite{cbpr} by diagonalizing 
the Cu$_{5}$O$_{4}$ cluster with increasing number 
$n_{h}$ of holes. We wish to demonstrate that the power of the 
symmetry driven mechanism is such that  attractive interactions arise even
in small clusters with $n_{h}>4$, despite the high hole concentration.
One can proceed from the true hole vacuum and insert 
holes until the last two form a W=0 pair; if the interactions produce 
a bound state we conventionally speak of {\em hole pairing}. 
Alternatively, one can proceed from the true electron vacuum and insert 
electrons until the last two form a W=0 pair; if the interactions produce 
a bound state we conventionally speak of {\em electron pairing}. These 
two expressions simply mean that we get pairing by adding two holes 
(as in La$_{2}$CuO$_{4}$) or two electrons (as in (Nd,Ce)$_{2}$CuO$_{4}$), 
respectively\cite{note}. The physical point here is that electron pairs and hole pairs
are related by a charge conjugation symmetry and the very same basic mechanism or
diagram is operating in both cases. We find new instances of electron and hole
pairing, again with a binding energy of a few tenths of meV in the physical parameter 
space. We also find a case when a W=0 pair leads to a weak repulsion.
 Further, we demonstrate how the two
different symmetries (A$_{1}$ and B$_{2}$) of W=0 singlet pairs allowed by
the cluster are both necessary to produce the superconducting flux quantisation
phenomenon in this cluster.

One reason for considering clusters again, after much excellent 
work from several authors \cite{dagotto} and our own previous work on 
clusters and on the plane,  is that in this 
way we can test our canonical transformation approach against the results of exact 
diagonalisation.
Another reason is that we wish to explore the relation of the flux 
quantisation phenomenon to the symmetry Group in the presence of the 
vector potential, which breaks the translational symmetry.
Our main questions are: can pairing be reliably predicted by studying 
the behavior of the system at weak coupling? Is the 
superconducting flux 
quantisation property exclusive of small clusters, or is it  a general 
consequence of symmetry?

\section{ ONE-BODY ENERGY LEVELS OF THE symmetric 9-site CLUSTER}

 In the hole picture,
the one-body energy levels of $Cu_{5}O_{4}$ are those displayed in Table \ref{livelli}. 
Here, the hole vacuum is a state with no holes at all.

Two levels are triply degenerate, comprising twice degenerate states of e(x,y) 
symmetry and states belonging to $b_{1}$; this {\em accidental}  degeneracy is due to 
the fact that in this small cluster any permutation of the four 
$Cu-O$ units bound to the central Cu is a symmetry; therefore, the full symmetry Group 
of the cluster is $S_{4}$, which has $C_{4v}$ as a subgroup, and 
admits degeneracy 3. Since 
this property does not extend to the plane, we continue using the 
irreducible representations (IRREPS) of $C_{4v}$ anyhow. 

In the electron picture, the levels are met in reverse order, but the 
sequence of symmetry labels remains the same. Thus, one 
notices that there is an approximate electron-hole symmetry, or 
charge conjugation symmetry, in this model.

\section{ W=0 PAIRS}

Both in the full plane and in clusters, the W=0 pairs are due to the 
symmetry, but there are some differences between the two cases, that 
we wish to stress in this Section. Let us first review the theory for 
the plane\cite{csb}. Omitting the
band indices, we shall mean 
\begin{equation}
\left| d[k] \right\rangle=\left\| k_{+},-k_{-}\right\| =c_{k,+}^{\dagger}c_{-k,-}^{\dagger}|vac>  
\label{kpairs}
\end{equation}
to be a two-hole determinantal state derived from the Bloch 
eigenfunctions ($|vac>$ is the true hole vacuum).

The point symmetry Group of the Cu-O plane is $C_{4v}$. We introduce the determinants $Rd[k]=d[Rk],
R\in C_{4v}$ , and the projected states 
\begin{equation}
\Phi _{\eta }\left[ k\right] =\frac{1}{\sqrt{8}}{\sum_{R\in C_{4v}}}\chi
^{\left( \eta \right) }\left( R\right) \left|Rd[k]\right\rangle  \label{project}
\end{equation}
where $\chi ^{\left( \eta \right) }(R)$ is the character of the operation $R$
in the Irrep $\eta $. In the non-degenerate
Irreps, the operations that produce opposite $Rk$ have the same character,
and the corresponding projections lead to singlets. Let $R_{i},i=1,..8$
denote the operations of $C_{4v}$ and $k,k^{\prime} $ any two points in the
Brillouin Zone (BZ). Consider any two-body operator $\hat{O}$, which is
symmetric ($R_{i}^{\dagger}\hat{O}R_{i}=\hat{O}$), and the matrix with elements $%
O_{i,j}=<d[k]|R_{i}^{\dagger}\hat{O}R_{j}|d[k^{\prime }]>$, where $k$ and $k^{\prime }$ may be taken to be in the same or in different bands. This
matrix is diagonal on the basis of symmetry projected states, with
eigenvalues 
\begin{equation}
O\left( \eta ,k,k^{\prime }\right) ={\sum_{R}}\chi ^{\left( \eta
\right) }\left( R\right) O_{R}\left( k,k^{\prime }\right)   \label{5}
\end{equation}
where 
\begin{equation}
O_{R}\left( k,k^{\prime }\right) =\left\langle d[k]|\hat{O}
|Rd[k^{\prime }]\right\rangle .  \label{projected}
\end{equation}
Thus, omitting the $k$, $k^{\prime }$ arguments, we get in particular 
\begin{eqnarray}
O\left( ^{1}A_{2}\right)  &=&O_{E}+O_{C_{2}}+O%
_{C_{4}}+O_{C_{4}^{3}}  \nonumber \\
&&-O_{\sigma _{x}}-O_{\sigma _{y}}-O_{\sigma _{1}^{\prime
}}-O_{\sigma _{2}^{\prime }}  \label{7}
\end{eqnarray}
\begin{eqnarray}
O\left( ^{1}B_{2}\right)  &=&O_{E}+O_{C_{2}}-O%
_{C_{4}}-O_{C_{4}^{3}}  \nonumber \\
&&-O_{\sigma _{x}}-O_{\sigma _{y}}+O_{\sigma _{1}^{\prime
}}+O_{\sigma _{2}^{\prime }}  \label{8}
\end{eqnarray}
If $\hat{O}$ is identified with W, since $W_{E}=W_{C_{2}}=W_{\sigma
_{x}}=W_{\sigma _{y}}$ and $W_{C_{4}}=W_{C_{4^{3}}}=W_{\sigma _{1}^{\prime
}}=W_{\sigma _{2}^{\prime }}$, one finds $W\left( ^{1}A_{2}\right) =W\left(
^{1}B_{2}\right) =0$ . These are W=0 pairs, like those studied 
previously \cite{cbpr} in clusters. In the full plane, however, W=0 pairs are 
obtained from holes at the Fermi
level for any filling.

Small clusters like $Cu_{5}O_{4}$ allow a nice 
 illustration of the theory because they also allow  W=0 2-body 
 solutions. This property is a consequence of their full $C_{4v}$ 
 symmetry around the central Cu. However, there are no Bloch states 
 in a finite cluster with open boundary conditions, and the W=0 singlet pairs come out differently. 
 First, we may consider  the orbitals of (x,y) symmetry in Table \ref{livelli}, 
 and form 2-hole determinants
 \begin{eqnarray}
 d[x,y]=\left\| x_{+},y_{-}\right\| 
 =c_{x+}^{\dagger}c_{y-}^{\dagger}|vac>\nonumber\\
 d[y,x]=\left\| y_{+},x_{-}\right\| =c_{y+}^{\dagger}c_{x-}^{\dagger}|vac>;
 \label{d[x,y]}
 \end{eqnarray}
they are eigenstates of $H_{0}$ and have the W=0 property, since the 
amplitude of double occupation of any site is 0.  Unlike the case of the full 
plane, no projection like that performed in Equation (\ref{project}) is 
necessary here to get the property. In $Cu_{5}O_{4}$, 
there are two sets of $(x,y)$ states, so the $x$ and $y$ of the above 
equation may belong to the same or to different sets. If the $x$ and 
$y$ states are taken from the same set,  the singlet 
\begin{equation}
\psi_{1}(^{1}B_{2})=\frac{d[x,y]+d[y,x]}{\sqrt{2}}
\label{pairs}
\end{equation}
 is an eigenstate of the kinetic energy and of $W$, belongs to 
the W=0 eigenvalue and to $^{1}B_{2}$. W=0 pairs of this 
symmetry and of $^{1}A_{2}$ exist in the full plane as well.
If the $x$ and $y$ states are taken from the different sets, we denote 
one of the sets by a prime and consider two-hole determinants like 
$d[x,y^{\prime}]$; these are eigenstates of $H_{0}$ and are W=0 pairs, 
however they do not belong to any of the IRREPS of $C_{4v}$.  We can form singlet 
combinations with the W=0 property both in the $^{1}B_{2}$ and  
$^{1}A_{2}$ symmetry, namely,
\begin{equation}
\psi_{2}(^{1}B_{2})=\frac{d[x,y^{\prime}]+d[y,x^{\prime}]+d[x^{\prime},y]
+d[y^{\prime},x]}{2},
\label{pairs1b2}
\end{equation}
which belongs to $^{1}B_{2}$ and
\begin{equation}
\psi(^{1}A_{2})=\frac{d[x,y^{\prime}]-d[y,x^{\prime}]+d[x^{\prime},y]
-d[y^{\prime},x]}{2}
\label{pairs1a2}
\end{equation}
which belongs to $^{1}A_{2}$. In addition, there are also  $^{1}A_{1}$ 
W=0 pairs, using the degenerate $x,y$ and $b \equiv b_{1}$ orbitals. 
The two-hole states 
\begin{equation}
\frac{
\left\|x_{+}x_{-}\right\|+\left\|y_{+}y_{-}\right\|}
{\sqrt{2}} \equiv |x^{2}+y^{2}>
\end{equation}
and
\begin{equation}
\left\| b_{+}b_{-}\right\| \equiv |bb>
\end{equation}
 are a basis of degenerate eigenstates 
of $H_{0}$  having $^{1}$A$_{1}$ symmetry. Diagonalising  $W$ in this 
basis we get two-hole eigenstates  of $H$. The $2 \times 2$ matrix of 
W is: 
\begin{eqnarray}
\left| 
\begin{array}{cc}
<bb|W|bb> & <bb|W|x^{2}+y^{2}> \\ 
<bb|W|x^{2}+y^{2}> & <x^{2}+y^{2}|W|x^{2}+y^{2}>
\end{array}
\right| =  \nonumber \\
\left| 
\begin{array}{cc}
\frac{U_{p}}{4} & \frac{U_{p}}{2\sqrt{2}} \\ 
\frac{U_{p}}{2\sqrt{2}} & \frac{U_{p}}{2}.
\end{array}
\right|  \label{a1pairs}
\end{eqnarray}
 The lowest
eigenvalue is 0 and the W=0 pair is 
\begin{equation}
\psi(^{1}A_{1}) =-\sqrt{\frac{2}{3}}|bb\rangle +
\sqrt{\frac{1}{3}}|x^{2}+y^{2}>.
\end{equation}
 This type of W=0 pairs does not exist in the full plane.
 The upper eigenvalue is $\frac{3U_{p}}{4}$ and the
eigenfunction $\sqrt{\frac{1}{3}}|bb\rangle +\sqrt{\frac{2}{3}}%
|x^{2}+y^{2}>$ is strongly affected by the on-site repulsion.

Now consider the $Cu_{5}O_{4}$ cluster in the non-interacting limit  with 2 holes,
which sit the lowest level of $a_{1}$ symmetry that we denote $a$ for 
short. Let this be the new vacuum state $|0>$. Adding 2 
holes, we partially fill the next degenerate levels, which can give 
rise to W=0 pairs. Starting with the above defined two-hole $d$ 
determinants, we can form 4-hole determinantal states, denoted 
by a capital D, like for example
\begin{eqnarray}
 D[x,y]=c_{x,+}^{\dagger}c_{y,-}^{\dagger}|0>\nonumber\\
 \equiv \left\| x_{+}a_{+},y_{-}a_{-}\right\|;
 \label{d[xa,ya]}
 \end{eqnarray}
in keeping with the notation of Ref.\cite{csb}, below we shall denote such
configurations with the background $a$ orbital occupied 
for both spins as $m$ states. We shall also need the other 4-hole 
determinantal states, namely, the $\alpha$ states, 
in which one of the background $a$ spin-orbitals is not occupied by holes, and 
the $\beta$ states in which both background $a$ spin-orbitals are missing.
For the $Cu_{5}O_{4}$ cluster in the non-interacting limit a similar 
situation is realised if the new vacuum state is taken with 10 holes,
filling the lowest levels according to the {\em aufbau} principle.  Adding 2 
holes, again we partially fill the next degenerate levels, which can give 
rise to W=0 pairs. A similar definition of $m,\alpha$ and $\beta$ is 
possible, and more excited configurations also exist. 

The matrix elements of W in this model have no exchange terms, since 
only holes of opposite spin can interact. The diagonal elements $W_{m,m}$ 
can be expressed in terms of orbitals $p,q,r,s$ by the two-hole integrals 
\begin{equation}
	W(p,q,r,s)=\sum\limits_{i}U_{i}p^{*}(i)q^{*}(i)r(i)s(i).
\end{equation}
  For example, the $m$ state of Equation(\ref{d[xa,ya]}) yields
\begin{equation}
W_{m,m}=W(x,a,x,a)+W(y,a,y,a)+W(a,a,a,a).
\end{equation}
We note that this is the Hartree-Fock interaction. The last term of the expression 
refers to the interaction between the background particles in the $a$ 
spin-orbitals and is the same for all the $m$ states, and the rest 
brings out single-particle corrections to the energy of the orbitals 
and could be readsorbed in the definition of $H_{0}$. The important 
point is that no term contains both $x$ and $y$; no direct 
interaction between the two added particles  exists, because of the W=0 property.

The matrix element of the two-body operator W between determinants 
which differ by two spin-orbitals are given by the well-known rule
\begin{eqnarray}
 \left\langle\left\| k_{+},k_{-},u_{1}\ldots u_{n}\right\| W \left\|
 k^{\prime}_{+},k^{\prime}_{-},u_{1}\ldots u_{n}\right\|\right\rangle\nonumber\\
 =\left\langle\left\| k_{+},k_{-}\right\|W \left\|
 k^{\prime}_{+},k^{\prime}_{-}\right\|\right\rangle\,
 \label{2body}
 \end{eqnarray}
where $k$ is different from $k^{\prime}$ while $u_{1}\ldots u_{n}$ is a 
sequence of occupied spin-orbitals. Using Equation (\ref{projected}), 
one finds that in the full plane, the matrix elements between different $m$ states 
$W_{m,m^{\prime}}$ vanish. In the cluster, this is not true. Let $x$ 
denote the state of $x$ symmetry taken from the lower $e$ degenerate 
level, and $y^{\prime}$ denote the state of $y$ symmetry taken from the 
upper $e$ degenerate level. Then, the $m$ states involving these 
orbitals are coupled by $W$ to those involving $x$ and $y$, and to those involving 
$x^{\prime}$ and $y^{\prime}$. Such matrix elements are forbidden in 
the full plane by momentum conservation, but exist in finite systems 
with open boundary conditions, having no translational invariance. 
They couple $m$ states belonging to different eigenvalues of 
$H_{0}$.

\section{THE EFFECTIVE INTERACTION}
 We need a rigorous definition of the effective interaction between 
 two holes in many-body systems, and this requires a careful analysis. 
 Actually, we shall use two alternative definitions, one of which is 
 suitable for numerical exact diagonalisation work, while the other 
 one is much more microscopic and analytical. Therefore, we have to 
 show that these two definitions essentially agree and lead to the same 
 physical conclusions. That will emerge from the analytical treatment 
 of the present Section and from the numerical results of Section 6.

  \subsection{First definition: $\Delta$}
 When working by exact diagonalisation, we consider a cluster with $n_{h}$ holes;
 its interacting ground state energy $E_{h}(n_{h}$), obtained with 
 the Hamiltonian of Equations (\ref{h},\ref{h0},\ref{w}), is 
 referenced to the hole vacuum for any $n_{h}$. In terms of  these 
 eigenvalues we define, following references (\cite{hirsh}, \cite{bal} ),
\begin{equation}
\Delta_{h} (n_{h})=E_{h}(n_{h})+E_{h}(n_{h}-2)-2E_{h}(n_{h}-1). 
\label{delta}
\end{equation}
 $\Delta_{h} (n_{h})$ is one definition of the pairing energy. 
 This definition is simple, but requires computing the eigenvalues 
with great accuracy, and has several drawbacks. It says nothing about 
the dynamics which leads to pairing. Moreover, generally a negative $\Delta$ does not
unambiguously imply
pairing, and further problems arise \cite{tinka}  since the 
above definition depends on the comparison of systems with different $n_{h}$. 

However, the application of Equation (\ref{delta}) is safe in the 
specific case when the last two holes are in a W=0 state; in Ref.\cite{cbpr}, we have shown 
that in this case $\Delta$ really coincides with the ground state expectation 
 value of the effective interaction, at 
least at weak coupling; if the interaction is attractive 
and produces a bound state, $\Delta_{h} (n_{h})$ is negative and  $|\Delta _{h}(n_{h})|$  
is the binding energy. These results were obtained by analysing exact 
diagonalisation results for clusters with $n_{h}=4$ by lowest-order 
perturbation theory. In the present paper, we wish to extend those results 
to larger $n_{h}$ by exact diagonalisation and a more powerful analytical 
method.
 
  \subsection{Second definition: $W_{eff}$}
  The alternative definition is intrinsic to the $n_{h}$ holes system 
and much more transparent. We achieve it by a canonical 
trasformation that determines the effective two-body Hamiltonian 
$\tilde{H}$ from 
the many-body H of Equation (\ref{h}). We set up the Schr\"{o}dinger 
equation for the ground state of the cluster with $n_{h}$ holes, namely
\begin{equation}
H |\Psi _{0}\rangle  = E_{0} |\Psi _{0}\rangle . \label{mb}
\end{equation}
Here, $E_{0} \equiv E_{h}(n_{h})$. We take the ground state configuration of the noninteracting 
$n_{h}-2$ system as our vacuum state (the non-interacting Fermi {\em sphere}).
 The exact $|\Psi _{0}>$ can be expanded in terms of excitations over the vacuum: 
 \begin{equation}
 |\Psi _{0}>={\sum_{m}}a_{m}|m>+{\sum_{\alpha }}b_{\alpha }|\alpha >+{\
 \sum_{\beta }}c_{\beta }|\beta >+....  \label{psi0}
 \end{equation}
 here m runs over pair states, $\alpha $ over 4-body states ($2$ holes and $1$
 e-h pair), $\beta $ over 6-body ones ($2$ holes and $2$ e-h pairs). 
 In $Cu_{5}O_{4}$ with 4 holes, the vacuum is the $a_{1}^{2}$ configuration 
 and  the expansion terminates with the $\beta$ states; it terminates anyhow in any finite system, after a 
 finite number of terms, so there are no convergence problems.
 Next, we consider the effects of the operators on the terms of $|\Psi _{0}>$. We write:
\begin{equation}
H_{0}|m>=E_{m}|m>,\;H_{0}|\alpha >=E_{\alpha }|\alpha >,...  \label{h0m}
\end{equation}
and since $W$ can create or destroy up to 2 e-h pairs,
\begin{eqnarray}
W|m>={\sum_{m^{\prime }}}W_{m^{\prime },m}|m^{\prime }>+{\sum_{\alpha }}%
|\alpha >W_{\alpha ,m}  \nonumber \\
+{\ \sum_{\beta }}|\beta >W_{\beta ,m}.  \label{wm}
\end{eqnarray}
For clarity let us
first write the equations that include explicitly up to 6-body states; then
we have 
\begin{eqnarray}
W|\alpha >={\sum_{m}}|m>W_{m,\alpha }+{\sum_{\alpha ^{\prime }}}|\alpha
^{\prime }>W_{\alpha ^{\prime },\alpha }  \nonumber \\
+{\sum_{\beta }}|\beta >W_{\beta ,\alpha }  \label{walfa}
\end{eqnarray}
where scattering between 4-body states is allowed by the second term, and

\begin{eqnarray}
W|\beta >={\sum_{m^{\prime }}}\left| m^{\prime }\right\rangle W_{m^{\prime
},\beta }+{\sum_{\alpha }}\left| \alpha \right\rangle W_{\alpha ,\beta } 
\nonumber \\
+{\sum_{\beta^{\prime } }}\left|\beta^{\prime } \right\rangle W_{
\beta^{\prime } ,\beta }  \label{wbeta}
\end{eqnarray}
 The Schr\"{o}dinger equation (\ref{mb}) yields equations for the
coefficients $a$,$b$ and $c$ 
\begin{eqnarray}
\left( E_{m}-E_{0}\right) a_{m}  \nonumber \\
+{\sum_{m^{\prime }}}a_{m^{\prime }}W_{m,m^{\prime }}+{\sum_{\alpha }}
b_{\alpha }W_{m,\alpha }+{\sum_{\beta }}c_{\beta }W_{m,\beta } =0
\label{coef1}
\end{eqnarray}

\begin{eqnarray}
\left( E_{\alpha }-E_{0}\right) b_{\alpha }  \nonumber \\
+{\sum_{m^{\prime }}}a_{m^{\prime }}W_{\alpha,m^{\prime }}+{\sum_{\alpha
^{\prime }}}b_{\alpha ^{\prime }}W_{\alpha ,\alpha ^{\prime }}+{\sum_{\beta }
}c_{\beta }W_{\alpha ,\beta } =0  \label{coef2}
\end{eqnarray}

\begin{eqnarray}
\left( E_{\beta }-E_{0}\right) c_{\beta }    \nonumber \\
+{\sum_{m^{\prime }}}a_{m^{\prime
}}W_{\beta ,m^{\prime }}+{\sum_{\alpha ^{\prime }}}b_{\alpha ^{\prime
}}W_{\beta ,\alpha ^{\prime }}+{\sum_{\beta ^{\prime }}}c_{\beta^{\prime} }
W_{\beta ,\beta^{\prime}}=0  \label{coef3}
\end{eqnarray}
where $E_{0}$ is the interacting ground state energy.
In principle, the $W_{\beta^{\prime } ,\beta }$ term can be eliminated by
taking linear combinations of the complete set of $\beta $ states. 
The complete set of $\beta $ states can be chosen in such a way that
\begin{equation}	
(H_{0}+W)_{\beta,\beta^{\prime}}=E^{\prime}_{\beta}\delta(\beta,\beta^{\prime})
\label{diag1}
\end{equation}
with this choice, the $W_{\beta^{\prime } ,\beta }$ terms are 
removed, while $E^{\prime}_{\beta}$ replaces the noninteracting 
eigenvalue $E_{\beta}$. In other terms, we get a self-energy correction to
$E_{\beta }$ and a mixing of the vertices, without altering the structure of the
equations. Then, we may rewrite 
Equation (\ref{coef3}) in the simpler form

\begin{equation}
\left( E^{\prime}_{\beta}-E_{0}\right) c_{\beta }+{\sum_{m^{\prime }}}a_{m^{\prime
}}W_{\beta ,m^{\prime }}+{\sum_{\alpha ^{\prime }}}b_{\alpha ^{\prime
}}W_{\beta ,\alpha ^{\prime }}=0. \label{coef4}
\end{equation}
 Now, we exactly decouple the
6-body states by solving the equation (\ref{coef4}) for $c_{\beta }$ and substituting into
(\ref{coef1},\ref{coef2}), getting:

\begin{eqnarray}
\left( E_{m}-E_{0}\right) a_{m}+{\sum_{m^{\prime }}}a_{m^{\prime }}\left[
W_{m,m^{\prime }}+{\sum_{\beta }}\frac{W_{m,\beta }W_{\beta ,m^{\prime }}}{
E_{0}-E^{\prime}_{\beta}}\right]  \nonumber \\
+{\sum_{\alpha }}b_{\alpha }\left[ W_{m,\alpha }+{\sum_{\beta }}\frac{
W_{m,\beta }W_{\beta ,\alpha }}{E_{0}-E^{\prime}_{\beta}}\right] =0  
\label{malfa1}
\end{eqnarray}

\begin{eqnarray}
\left( E_{\alpha }-E_{0}\right) b_{\alpha }+{\sum_{m^{\prime }}}a_{m^{\prime
}}\left[ W_{\alpha,m^{\prime }}+{\sum_{\beta }}\frac{W_{\alpha,\beta }W_{\beta ,m ^{\prime }}}{E_{0}-E^{\prime}_{\beta}}\right]  \nonumber \\
+{\sum_{\alpha ^{\prime }}}b_{\alpha ^{\prime }}\left[ W_{\alpha ,\alpha
^{\prime }}+{\sum_{\beta }}\frac{W_{\alpha
,\beta }W_{\beta ,\alpha ^{\prime }}}{E_{0}-E^{\prime}_{\beta}}\right] =0  
\label{malfa2}
\end{eqnarray}

Introducing  renormalised interactions $W^{\prime}$, we may rewrite 
these equations in the form
\begin{equation}
\left( E_{m}-E_{0}\right) a_{m}+{\sum_{m^{\prime }}}a_{m^{\prime
}}W_{m,m^{\prime }}^{\prime }+{\sum_{\alpha }}b_{\alpha }W_{m,\alpha 
}^{\prime }=0  \label{malfa3}
\end{equation}

\begin{equation}
\left( E_{\alpha }-E_{0}\right) b_{\alpha }+{\sum_{m^{\prime }}}a_{m^{\prime
}}W_{\alpha ,m^{\prime }}^{\prime }+{\sum_{\alpha^{\prime }}}b_{\alpha^{\prime
}}W_{\alpha ,\alpha^{\prime }}^{\prime }=0  \label{malfa4}
\end{equation}

If in Equations (\ref{coef1},\ref{coef2}) we drop the terms 
involving the $\beta$ states, they reduce to the same form as 
Equations (  \ref{malfa3},\ref{malfa4}), except that in the latter equations 
some quantities are renormalised. In other terms, the r\^{o}le of 6-body states is just
to renormalize the
interaction in the equations for the 2-body and 4-body ones, and for the rest they may be
forgotten about. If $E_{0}$ is outside the continuum of excitations, as we
shall show below, the corrections are finite, and experience with clusters
suggests that they are small. Had we included 8-body excitations, we could
have eliminated them by solving the system for their coefficients and
substituting, thus reducing to the above problem with further
renormalizations. This is a recursion method to perform the full 
canonical transformation; it applies to all the higher order
interactions, and we can recast our problem as if only $2-$ and 4-body states
existed.

Again, the $W^{\prime}_{\alpha^{\prime } ,\alpha }$ term can be eliminated 
from Equation (\ref{malfa4})
by taking linear combinations of the $\alpha $ states. This is 
achieved by choosing the complete set of $\alpha $ states in such a way that
\begin{equation}	
(H_{0}+W^{\prime})_{\alpha,\alpha^{\prime}}=E^{\prime}_{\alpha}\delta(\alpha,\alpha^{\prime}).
\label{diag2}
\end{equation}
With this choice, the $W_{\alpha ,\alpha^{\prime } }^{\prime }$ terms are 
removed, while $E^{\prime}_{\alpha}$ replaces the noninteracting 
eigenvalue $E_{\alpha}$. In other terms, we get a self-energy correction to
$E_{\alpha }$ and a mixing of the vertices, without altering the structure of the
equations. Now Equation (\ref{malfa4}) becomes

\begin{equation}
\left( E_{\alpha }^{\prime }-E_{0}\right) b_{\alpha }+{\sum_{m^{\prime }}}a_{m^{\prime
}}W_{\alpha ,m^{\prime }}^{\prime }=0  \label{malfa5}
\end{equation}
Solving Equation (\ref{malfa5}) for $b_{\alpha }$ and substituting in 
Equation (\ref{malfa3})  we exactly decouple the 4-body states as well.
The eigenvalue problem is now 
\begin{equation}
\left( E_{0}-E_{m}\right) a_{m}=\sum_{m^{\prime}} a_{m^{\prime }}
\left\langle m|S[E_{0}]|m^{\prime }\right\rangle  , 
\label{schro}
\end{equation}
where

\begin{equation}
\left\langle m|S\left[ E_{0}\right] |m^{\prime }\right\rangle =W_{m,m^{\prime }}^{\prime }+
{\sum_{\alpha
}}\frac{<m|W^{\prime }|\alpha ><\alpha |W^{\prime }|m^{\prime }>}{E_{0}-E_{\alpha }^{\prime }}. 
\label{eiv}
\end{equation}
We introduce the diagonal elements of the $\alpha$ summation:
\begin{equation}
F_{m,m}={\sum_{\alpha
}}\frac{<m|W^{\prime }|\alpha ><\alpha |W^{\prime }|m>}{E_{0}-E_{\alpha }^{\prime }};
\end{equation}
 then, Equation (\ref{schro}) becomes
\begin{eqnarray}
E_{0}a_{m}=(E_{m}+W^{\prime}_{m,m}+F_{m,m})a_{m}+\nonumber\\ 
{\sum_{m^{\prime}\neq m}}a_{m^{\prime}}\left\langle m|W_{eff}|m^{\prime }\right\rangle
\label{es}
\end{eqnarray}
where for $m\neq m^{\prime}$
\begin{equation}
\left\langle m\left |W_{eff} \right|m^{\prime }\right \rangle 
=W^{\prime}_{m,m^{\prime}}+{\sum_{\alpha
}}
\frac{
\langle m|W^{\prime }|\alpha \rangle\langle \alpha |W^{\prime }|m^{\prime }
\rangle}
{E_{0}-E_{\alpha }^{\prime }}.
\label{weff}
\end{equation}
The $W^{\prime}_{m,m^{\prime}}$ term does not arise in Ref. \cite{csb} 
because in the full plane it vanishes by momentum conservation.

Equations (\ref{es},\ref{weff}) determine the amplitudes $a_{m}$ of the $m$ 
states in the $n_{h}$-hole state and the ground state eigenvalue
$E_{0}$ relative to the hole vacuum. Their solution, inserted in 
Equation (\ref{malfa5}) yields the coefficients $b_{\alpha}$ and we 
could proceed with the full calculation of $\Psi_{0}$; this appears 
to be  hard for a large system. However, our 
task here is   to find the effective two-body Hamiltonianian; this is much less expensive.

Indeed, Equation (\ref{es}) is of the form of a Schr\"{o}dinger 
equation with eigenvalue $E_{0}$ for pairs with effective interaction 
$W_{eff}$.
Then we may interpret $a_{m}$ as the wave function of the {\em dressed pair}, which is acted upon by an
effective Hamiltonian $\tilde{H}$. The change from the full many-body $H$ to $\tilde{H}$ is a
canonical transformation which holds to all orders.  $W_{eff}$ is the effective interaction between
dressed holes, while $F$ is a
forward scattering operator, which
accounts for the self-energy corrections of the one-body propagators: it is
evident from (\ref{es}) that it just redefines  $E_{m}^{\prime }$. Also in Cooper 
theory\cite{kn:kittel} one meets electron-phonon self-energy terms, 
which do not contribute to the effective interaction. The basic spin-flip diagram responsible
for $W_{eff}$ had been identified before \cite{cbpr}. 
Any other pairing mechanism not considered here, like off-site
interactions, inter-planar coupling and phonons, can be included as 
an extra contribution to $W_{m^{\prime },m}^{\prime }$ which  just adds to $W_{eff}$.

This way of 
looking at Equation (\ref{es}) is perfectly consistent, despite the 
presence of the many-body eigenvalue $E_{0}$, because we are not 
compelled to reference the energy eigenvalues to the hole vacuum.  We note that if we shift
$H_{0}$ by an arbitrary constant $\Delta E$ in Equation (\ref{h0m}), by setting
\begin{equation}
H^{\prime}_{0}=H_{0}-\Delta E  \label{shift}
\end{equation}
the same shift applies to the 
eigenvalues $E_{m},E_{\alpha },E_{\beta }$ and so on, and also to the 
renormalised quantities like $E^{\prime}_{\alpha 
},E^{\prime}_{\beta }$. Therefore, the effective interaction 
$W_{eff}$ of Equation (\ref{weff}) and the $F$ matrix elements are unaffected 
by the shift. Thus we can reference $E_{0}$ to a new energy origin 
by shifting the diagonal terms in Equation(\ref{es}) without changing 
the off-diagonal terms. Since we wish to regard Equation (\ref{es}) 
as a Cooper-like equation for the pair, it is natural to set $\Delta E$ 
equal to the interacting ground state energy eigenvalue for the 
$n_{h}-2$ hole system, relative to the hole vacuum,
\begin{equation}
\Delta E=E_{h}(n_{h}-2)
\label{shift2}
\end{equation}
This quantity is 
obtained by diagonalising the cluster Hamiltonian with $n_{h}-2$ holes. 
  In Ref.\cite{csb}, dealing with the infinite plane, this was our choice. 
  The energy of two independent holes, relative to the  $n_{h}-2$ 
  background, is $2E_{F}$, where $E_{F}$ is the Fermi energy; when the 
  effective interaction is accounted for, the energy of the two bound 
  holes is $2E_{F}+\Delta$, where $|\Delta|$ is the binding energy.

 Up to this point, the treatment is exact. However, we can make an 
 easy use of Equation (\ref{es}) if we can neglect the 
 renormalisations in Equation (\ref{weff}), setting $W^{\prime} 
 \rightarrow W$ and $E_{\alpha}^{\prime} 
 \rightarrow E_{\alpha}$, which is fully justified 
 in the weak coupling case.  This is the approximation that we 
 proposed in Ref (\cite{csb}) and that we want to test in the present 
 paper. In fact, if we are primarily interested in the 
symmetry of the ground state, and in the presence or absence of 
pairing, we can get these results without a large 
computational effort. We exemplify the procedure for $Cu_{5}O_{4}$ in the $n_{h}=4$ case.
Two degenerate $m$ states are lowest in the non-interacting limit, 
namely, the configuration $m=D[x,y]$ of Equation (\ref{d[xa,ya]}) and 
$m^{\prime}\equiv D[y,x]$, where the $x,y$ orbitals belong to the lower $e$ level;
the $\psi(^{1}A_{1})$ state also is 
degenerate with $m,m^{\prime}$, but by symmetry $W$ cannot mix it to 
them.
  As already noted, $m$ and $m^{\prime}$ do not interact through 
the $W_{m,m^{\prime}}$ term. To calculate $W_{eff}$, we rewrite $W$  
(Equation (\ref{w})) in 
the orbital representation, with
\begin{equation}
c_{i}^{\dagger}=\sum_{\nu}^{orb}\langle i|\nu \rangle c_{\nu}^{\dagger}
\end{equation}
where $\nu$ runs over all the orbitals, obtaining
\begin{equation}
W=\sum_{\mu \nu \rho \sigma}^{orb} W(\mu,\nu,\rho,\sigma)
c_{\mu,+}^{\dagger}c_{\nu-}^{\dagger} c_{\sigma,-}c_{\rho,+}.
\end{equation}
The pair $(\rho_{+},\sigma_{-})$ which is annihilated may correspond 
to $(y_{+},x_{-})$, $(y_{+},a_{-})$, $(a_{+},x_{-})$, $(a_{+},a_{-})$.
The first choice gives nothing since it corresponds to a W=0 pair; the 
last choice yields a $\beta$ state.  To lowest order, only the 
$\alpha$ states contribute, involving the excitation of either the 
$a_{+}$ or the $a_{-}$ hole. Many of the $W$ matrix elements vanish by 
symmetry; we are going to neglect those connecting to excited $x',y'$ 
orbitals, which occur at higher energies, because we are considering 
weak coupling.

Considering the contribution of $(y_{+},a_{-})$, one finds that the only  $\alpha$ 
states coupled to  $D[y,x]$  by $W$ are those of the form $|\mu a y x|\equiv |\mu_{+} 
a_{+} y_{-} x_{-}|$, in which  the hole in $a_{-}$ is promoted to 
$y_{-}$ while $y_{+}$ is scattered into $\mu_{+}$. Therefore,
\begin{equation}
E_{\alpha}=\varepsilon_{a}+\varepsilon_{\mu}+
\varepsilon_{x}+\varepsilon_{y},
\end{equation}
where $\varepsilon_{y}=\varepsilon_{x}$. One finds
\begin{equation}
\langle \left|\mu a y x\right |W\left |y a x a\right | \rangle =-W(y,a,\mu,y)
\end{equation}
and
\begin{equation}
\langle \left |x a y a \right|W \left |\mu a y x\right| \rangle =W(x,a,\mu,x).
\end{equation}
Therefore, taking into account that each of the two background $a$ holes can be 
promoted and this brings a factor of 2, using (\ref{es}) we obtain
\begin{equation}
\left\langle m\left |W_{eff} \right|m^{\prime }\right\rangle 
=-2{\sum_{\mu}}
\frac{
W(y,a,\mu,y)W(x,a,\mu,x)
}
{E_{0}-(\varepsilon_{a}+\varepsilon_{\mu}+
2 \varepsilon_{x})}. 
\label{weff2}
\end{equation}
Since $W(y,a,x,y)=W(x,a,y,x)=0$, the empty (of holes) states $\mu$  belonging to the $e$ representation yield 0. 
The empty orbitals that contribute are those of the $a_{1}$ symmetry, that will be denoted by $a^{\prime }$,
 and those of $b_{1}$ symmetry that we shall write $b$.

 The $a^{\prime }$ orbitals contribute to the repulsion, and the $b$ 
 orbitals to the attraction. Indeed, $W(x,a,a^{\prime },x)=W(y,a,a^{\prime 
 },y)$, and the contribution of the  states of $a_{1}$ symmetry is 
$ -2{\sum_{a^{\prime }}}
\frac{W(x,a,a^{\prime },x)^{2}}{E_{0}-(\varepsilon_{a}+\varepsilon_{a^{\prime}}+
2 \varepsilon_{x})}$; since $E_{\alpha} > E_{0}$ this is positive. On the other hand,
$W(x,a,b,x)=-W(y,a,b,y)$, since the orbitals of of $b_{1}$ symmetry 
change sign for a $\frac{\pi}{2}$ rotation. Therefore the 
contribution of the $b$ states is attractive. This is an example of the interference of opposite contributions to 
 $W_{eff}$, that we emphasised in Ref \cite{csb}.

Therefore, at this stage,  a self-consistent treatment
of Equation (\ref{es}) must be
sought, because $W_{eff}$ depends on the eigenvalue 
$E_{0}$. A straightforward recursion approach leads to a continued fraction 
solution, which has contributions from all orders of perturbation 
theory. 

  \subsection{Equivalence of the two definitions at weak coupling}
  
For weak coupling, however, a cruder but simpler  appoximation 
is justified: one calculates $W_{eff}$ 
neglecting all self-energy corrections, in such a way that $E_{0}$ in 
(\ref{weff2}) reduces to $2 \varepsilon_{x}+2 \varepsilon_{a}$; this is the lowest
(second-order) approximation $W_{eff}^{(2)}$.  
In the same, lowest-order approximation, the shift in Equation (\ref{shift2}), 
which recasts Equation (\ref{es}) as a two-body problem, in the Cooper-like
form, is $\Delta E=2 \varepsilon_{a}$; further, one considers only the 
mixing of the degenerate configurations $m=D[x,y]$  and 
$m^{\prime}\equiv D[y,x]$. In the resulting  $2\times 2$ problem, the diagonal 
entries are identical, and  the $W_{eff}^{(2)}$ interaction produces the 
off-diagonal elements, with the result that the 
singlet is stabilised by $|W_{eff}^{(2)}|$ and the triplet is destabilised 
by the same amount. Therefore \cite{nota2},  $\Delta (4)^{(2)}=W_{eff}^{(2)}$. One obtains for 
$\Delta(4)^{(2)} $ the following second-order expression: 
\begin{eqnarray}
\Delta(4)^{(2)} =\nonumber\\
-2\left[ \sum\limits_{b}\frac{W(a,b,x,x)^{2}}{(\epsilon
_{b}-\epsilon _{a})}-
\sum\limits_{a^{\prime }}\frac{W(a,a^{\prime },x,x)^{2}%
}{(\epsilon _{a^{\prime }}-\epsilon _{a})}\right]   \label{delta2}
\end{eqnarray}
where  the sums run only over the one-body states of $a$ and $b$ symmetry. This 
agrees with 
the result that we obtained earlier \cite{cbpr} from a diagramatic 
analysis of Equation (\ref{delta}). Thus, the two definitions of the 
effective interaction lead to the same result, at least in the weak 
coupling limit.

We stress again that the sign of $\Delta(4)^{(2)}$
is determined by the relative weights of the virtual excitations to the empty
states of different point symmetry. In general, $W_{eff}$ can produce attraction or 
repulsion, depending on the hole concentration. This crude approximation 
will turn out to be sufficient for qualitative purposes, i.e, to 
predict when pairing occours.

\section{ W=0 PAIRS AND CHARGE CONJUGATION}
Consider the cluster with $n_{h}$ holes. As shown above, the interesting situation arises when 
$n_{h}$ is such that, filling the levels according to the
{\em aufbau} principle,the last two holes go to a degenerate level. 
Accordingly, 
we expect that $\Delta_{h} (n_{h})$ measures the effective interaction 
between the holes of the W=0 pair. In Ref.\cite{cbpr}, we have shown 
that this is the case at weak coupling. If the interaction is attractive 
and produces a bound 
state, $|\Delta _{h}(n_{h})|$  is the binding energy. This situation can be realised with $n_{h}=4$ in  highly symmetric Cu-O clusters
 containing up to 21 atoms\cite{cb1}. The last 2 holes then go to the 
 lowest level of $e$ symmetry.
 
 According to Table I, the Cu$_{5}$O$_{4}$ cluster has an 
 upper $e$ level, which is reached with 12 holes, so we are 
 interested in  $\Delta _{h}(4)$ and $\Delta _{h}(12)$. Moreover, we 
 can exploit the approximate electron-hole symmetry of the problem to 
 obtain two more interesting cases. The approximate symmetry consists 
 in the fact that the same sequence of symmetry labels is obtained by 
 reading Table \ref{livelli} from up to down and in reverse order. The reverse 
 order corresponds to adopting the electron picture and starting from 
 the electron vacuum. Going to the electron picture, the three-band
Hubbard Hamiltonian  becomes: 
\begin{eqnarray}
H &=&\sum_{i}(2\varepsilon _{i}+U_{i})-\sum_{i\sigma }(\varepsilon
_{i}+U_{i})a_{i\sigma }^{\dagger}a_{i\sigma }  \nonumber \\
&&-\sum_{<i,j>\sigma }t_{ij}a_{i\sigma }^{\dagger}a_{j\sigma
}+\sum_{i}U_{i}n_{i+}n_{i-},  \label{helectrons}
\end{eqnarray}
  W=0 {\em electron pairs} are 
obtained for $n_{e}$=4 and 12 electrons.  Letting now 
$E_{e}(n_{e}$) denote the ground state energy of the cluster with $n_{e}$ 
electrons, the effective interaction between the two electrons in the 
pair is measured by  
\begin{equation}
\Delta_{e} (n_{e})=E_{e}(n_{e})+E_{e}(n_{e}-2)-2E_{e}(n_{e}-1).
\end{equation}
Since the dimensionality of the one-body basis is 18,
\begin{eqnarray}
\Delta_{e} (4)=E_{e}(4)+E_{e}(2)-2E_{e}(3)=\nonumber \\
E_{h}(14)+E_{h}(16)-2E_{h}(15)=\Delta_{h} (16)
\end{eqnarray}

and, similarly, $\Delta_{e} (12)=\Delta_{h} (8)$. 

We recall that we speak of {\em electron pairs} when two added 
electrons partially occupy a degenerate state and of {\em hole pairs} 
when the same situation is reached by adding two holes; however the 
final situation is exactly the same. For example, consider the W=0 pair state of
Eq. (\ref{pairs}). One readily verifies that in a canonical transformation from
holes to electrons, putting $ a_{i\sigma }^{\dagger}$=$c_{i\sigma }$, the
two-hole state  becomes a two electron state of the same form.
Therefore the two-body W=0 state is invariant under charge 
conjugation, and if holes are paired, electrons are also paired.
In order to avoid switching all the time between the two equivalent pictures, 
below we discuss everything in terms of holes. Summarizing the results 
of the present Section, we can test  the effective interaction 
in Cu$_{5}$O$_{4}$ by calculating $\Delta_{h} (n_{h})$ with 
$n_{h}=$ 4,8,12 and 16.

\section{ NUMERICAL RESULTS AND DISCUSSION}

By an enhanced Lanczos routine, we computed the ground state of the Cu$_{5}$O$_{4}$
cluster with even numbers $n_{h}$ of holes and vanishing $z$ component of the total
spin;  the parameter values are specified in the Introduction. 
The maximum size of the matrices  (15,876) 
occurs for $n_{h}=8$; the $n_{h}=$12 and 16 cases are handled by 
transforming to the electron picture. The results for $\Delta_{h} (n_{h})$ are summarised in Table \ref{meV}. One 
sees that $\Delta_{h}$ for $n_{h}=$ 4,8,12 and 16 is much smaller in absolute 
value than for other fillings, as expected. This confirms that W=0 
pairs are involved. In particular, for $n_{h}=$ 4,8 and 16 
$\Delta_{h}(n_{h})<0$ and pairing occurs, while for $n_{h}=$12 a 
small repulsion prevails. When pairing is obtained, this means that 
the renormalisation of the parameters inherent in the canonical 
transformation does not have important consequences.  To see if the behaviour at $n_{h}=$12 is 
an exception, we have repeated the calculations with scaled 
$U$ values. Using $U_{p}=$ .06 eV and $U_{d}=$ .053 eV , which are 
$<<t$ and allow applying perturbation theory, we still get a 
positive result, namely $\Delta_{h}(12)$=0.0034 meV. Thus, even second-order 
perturbation theory would suffice 
to predict $\Delta_{h}(12)>0$ in this case.

For $n_{h}=4$ we have an analytic second-order result (Equation \ref{delta2})
and we can check its degree of validity  by comparing  with the exact
diagonalisation values of $\Delta $. Making
use of the above standard values of the parameters we calculate the 
relative error
$\delta =2\left| \frac{\Delta -\Delta^{(2)}}{\Delta +\Delta^{(2)}}\right| $. 
It turns out that $\delta \leq .07$  up to U/t $\approx $1. Thus, 
already the second-order approximation is remarkably accurate in 
estimating the effective interaction.
We conclude that our treatment based on the unrenormalised formula of 
Equation (\ref{weff}) correctly predicts the presence or absence of pairing, depending of 
the hole concentration, and even a simple, second-order approximation 
to it has a semi-quantitative accuracy when compared with exact results.

 The indications that we 
may draw from this Section are: i) our mechanism 
is operating for a wide range of hole concentrations and 
produces a much reduced interaction $|\Delta|$, ii) this does not imply pairing at all 
concentrations, iii) we can predict if there is pairing or repulsion 
in a particular case by our theory. The cluster approach, however, has several 
limitations, the main size 
effect being that W=0 pairs are possible at discrete values of the hole 
concentration.   In $Cu_{5}O_{4}$ with 4 holes, we are 
doping with one electron, but in other cases we are far from the 
physical concentrations.  However, we are not yet trying to make quantitative 
predictions, rather our point here is that of testing 
our approach against exact solutions, which is only feasible in small 
clusters.

\section{FLUX QUANTIZATION AND PAIR SYMMETRY}

If a magnetic field is confined to a hole in any material (flux tube) 
the flux $\phi$ is quantised in integer multiples of the fundamental quantum 
$\phi_{0} =\frac{hc}{e}$; a flux $\phi=\phi_{0}$ can be gauged away, 
and any physical property, for example the ground 
state energy, is a periodic functions of $\phi$ with period 
${\phi_{0}}$. 

Bulk superconductors quantize the flux through a hole 
in integer \underline{and} half-integer multiples of $\phi_{0}$, because 
the quasiparticles that screen the vector potential carry charge 
$2e$.   In finite systems the signature of 
superconductivity is  a ground state energy minimum at $\phi=0$ that is separated
by a barrier from a second minimum at  $\phi=\phi_{0}$/2.  With 
increasing the size of the system, the energy (or free energy, at 
finite temperature) barrier separating the two minima becomes 
macroscopic, 
and bulk superconductors can swallow up only an integer or 
half integer number of flux quanta. As emphasized by Canright and Girvin\cite{cg},
the flux dependence of the ground state energy is 
definitely a
most compelling way of testing for superconductivity, and the 
existence of the two minima separated by a barrier is  a strong indication of 
superconducting flux quantisation. 

In Ref. \cite{cg}, superconducting pairing was obtained by assuming a  
negative $U$; a {\em ribbon} shaped cluster was closed on itself with periodic 
boundary conditions along its length, and the flux was inserted in the 
hole. In the present problem, with a repulsive Hubbard model, the mechanism 
of attraction is driven by the $C_{4v}$ 
symmetry, and  cannot operate with  such an unsymmetric geometry.  The flux must be inserted in such a way
that the system is not distorted. 
On the other hand, we cannot make holes  in our small cluster because it would fall apart in 
disconnected pieces.  One should consider larger clusters like 
$Cu_{13}O_{36}$, which allow W=0 solutions for $n_{h}\geq 10$, 
however the number of configurations $>10^{12}$ is outside the scope 
of exact diagonalisation methods. 

  So, we keep the $Cu_{5}O_{4}$ cluster geometry, but modify its topology by adding 
  a small hopping $t_{d}$
between the external Cu's, in order to introduce a closed path around 
the centre, where screening currents can respond.  Each $t_{d}$ bond forms a closed triangular
loop with the central Cu at the vertex (see Figure I).  This geometry is a compromise, because the 
 magnetic field penetrates our small cluster; however, it lends 
 itself to an extension to the full plane, such that only the 4 central 
 plaquettes feel a magnetic field, and the rest of the plane only 
 experiences a vector potential (see below). Finally, we observe that 
 a flux of the order of a fluxon in a macroscopic system would be a 
 small perturbation; in the small cluster, however, the perturbation 
 is small only if the 
hopping integral $t_{d}$ is taken  small compared
to $t$. Numerically, the computations were performed with 
 $t_{d} =\pm 0.01$ eV.

We introduce a tube carrying flux $\phi $
inside each of the triangles formed in this way. Every bond collects  the 
Peierls phase $\frac{2 \pi i \int {\bf A}\cdot d{\bf r}}{\phi_{0}}$; 
by symmetry, $t$ is unaffected by the flux, while 
\begin{equation}
t_{d} \rightarrow t_{d}e^{\frac{2 \pi i \phi}{\phi_{0}}}  \label{td}
\end{equation}
for a clockwise path, and the complex conjugate expression  a counterclockwise path. 
 
 \subsection{Superconducting flux quantisation: numerical results}
 According to Table II, $\Delta_{h}(n_{h})$ is negative and pairing 
 results  at $\phi=0$ for $n_{h}$=4,8 and 16;
in all three cases we found that the ground state energy $E_{h}(n_{h}, \phi)$ as a function of 
$\phi$ has clearly separated minima at zero and half a flux quantum. 
Moreover, our criterion for pairing ($\Delta <0$) also leads us to a 
much more stringent criterion for superconducting flux quantisation 
than is drawn from the literature, since we need that both minima 
in the ground state versus flux curves also correspond to negative 
$\Delta$. This is a much clearer signature of
superconducting flux quantisation than the generally accepted presence 
of the two minima, because it implies that the superconductor remains 
a superconductor after swallowing up the half flux quantum. Therefore,
we computed $\Delta_{h}(n_{h}, \phi)$ in order to determine the flux dependence 
of the effective interaction. {\em When  $\Delta_{h}(n_{h}, 0)<0$, then
$\Delta_{h}(n_{h}, \frac{\phi_{0}}{2})$ is also negative.} For small 
enough $t_{d}$, the response function
\begin{equation}
R=\frac{\Delta_{h}(n_{h}, \phi)-\Delta_{h}(n_{h}, 0)}{|t_{d}|} 
\label{r}
\end{equation}
is an intrinsic property of the original cluster with $t_{d}=0$. 
In Figure II we show  $R$ for several $n_{h}$ values versus 
$\frac{\phi}{\phi_{0}}$. All the $R$ curves  have a local minimum at 
$\phi=0$, where they vanish;  $\phi=\phi_{0}$/4  is a maximum 
and a second minimum occurs at 
$\phi=\phi_{0}$/2; the $n_{h}=4$ curve is reduced by a factor of 3.  
The barrier gets lower with increasing $n_{h}$, but the same 
qualitative trend can be seen in all cases. The numerical data also 
show that changing the sign of $t_{d}$ produces a rigid shift of the 
$n_{h}$ curves by $\frac{\phi_{0}}{2}$ such that the two minima 
interchange their places.

In Figure II we also report 
the absolute value $|\left<\Psi_{0}(\phi)|\Psi_{0}(0)\right>|$ of 
the overlap between the the ground state eigenvectors in the presence 
and in the absence of the flux, for $n_{h}=4$ . It is clear that  
$|\left<\Psi_{0}(\frac{\phi_{0}}{2})|\Psi_{0}(0)\right>|=0$, and 
therefore the pairing state at zero flux and half fluxon are orthogonal.
There is a clear analogy with the BCS theory; in that case, the Cooper
wavefunction has $s$ symmetry  and the total magnetic quantum 
number of the pair vanishes in the absence of flux, but not at half a 
flux quantum\cite{g}. Similar results  for the overlap are obtained 
for the other $n_{h}$ values which correspond to partially filled 
shells.

 Our code automatically classifies the eigenvectors according to the 
IRREPS of $C_{4v}$. For positive $t_{d}$ the point symmetry of the 
ground state wavefunction changes from 
$^{1}$B$_{2}$($x^{2}-y^{2}$) at $\phi $=0 to $^{1}$A$_{1}$($x^{2}+y^{2}$) at
$\phi=\frac{\phi_{0}}{2}$. For negative $t_{d}$ the symmetry labels of 
the two minima are interchanged. For electron pairing, the 
symmetry of the states is the same as in the hole
case.

Since the vector potential lowers 
the symmetry, the eigenvectors cannot generally be classified according to the IRREPS of 
$C_{4v}$; however numerical data show that  at half fluxon, the symmetry is 
{\em dynamically} enhanced (see below).

\subsection{Group theory aspects of superconducting flux quantisation}

These findings are required by general symmetry principles. In the
absence of $t_{d}$, the full invariance Group of the cluster is $S_{4}$ 
and the interacting ground
state is degenerate, with $^{1}$A$_{1}$ and $^{1}$B$_{2}$ components. A
nonzero $t_{d}$ at $\phi=0$ reduces the symmetry to the $C_{4v}$ subgroup; 
it turns out that  with a positive $t_{d}$ the expectation value of 
the magnetic perturbation is negative on $^{1}B_{2}$ and positive on 
$^{1}A_{1}$; therefore  the ground state is $^{1}$B$_{2}$ at 
$t_{d}>0$ but changes symmetry if the sign of $t_{d}$ is reversed. Upon switching the vector
potential ${\bf A}$, the Cu-Cu hopping is complex and chiral, so the symmetry is 
lowered again from C$_{4v}$ to its  subgroup Z$_{4}$,
which contains only the rotations. Since $Z_{4}$  is abelian, there are no
degeneracies for a generic $\phi $, so  there are no W=0 pairs and
repulsion prevails. With increasing the flux  from 0, the ground state
energy increases to a maximum. Then it  decreases because, at $\phi 
=\frac{\phi_{0}}{2}$, the Cu-Cu hopping of Equation (\ref{td}) becomes $-t_{d}$, which is real; then
the full $C_{4v}$ symmetry is restored, resurrecting the W=0 pairs. The 
recovery of $C_{4v}$ at $\phi =\frac{\phi_{0}}{2}$ enables us to 
assign the eigenvectors to the IRREPs, as noted above. The change of symmetry of the pair is also readily 
understood: the perturbation caused by $t_{d}>0$ at $\phi=0$ becomes the 
opposite at half fluxon, so the $^{1}$A$_{1}$ state is lowest now. 
 The signature of superconducting pairing is not only the existence of a well defined 
second minimum at half flux quantum, but also the fact that it corresponds to a $\Delta <$0
situation, like at $\phi=0$.

This symmetry argument extends to the full plane. To see that, consider 
the pattern of Figure III. Here, the Cu sites are marked by X and the 
Oxygen sites by O; the black dots stand for tubes carrying flux 
$\phi$ each, symmetrically disposed around the central Cu. Varying 
$\phi$ by an integer multiple of $\phi_{0}$ corresponds to a gauge 
transformation leaving all the physical properties invariant. The arrows 
help to visualise a convenient choice of the gauge at general $\phi$. Namely, running 
along an oriented bond in the sense of the arrow, 
\begin{equation}
\int_{\rightarrow} {\bf A}\cdot d{\bf r}=\frac{\phi}{2};
\label{arrows}
\end{equation}
along the other Cu-O bonds, not marked in the 
Figure, $\int {\bf A}\cdot d{\bf r}=0$. One sees that in this way the flux through any 
closed path corresponds to the number of tubes surrounded by the path. 
The reflection operations of $C_{4v}$ are equivalent to 
$\phi\rightarrow -\phi$, reverse the directions of the 
arrows and for a generic $\phi$ the symmetry Group reduces to 
$Z_{4}$. However, at $\phi=\frac{\phi_{0}}{2}$ the reversal of the 
magnetic field in the tubes corresponds to a jump by $\phi_{0}$, and 
this is equivalent to a gauge transformation: this implies that the symmetry 
Group gets larger, the new symmetry operations being reflections 
supplemented by a gauge transformation. Indeed, it follows from 
Equation (\ref{arrows}) that the hopping parameter becomes $it$ along the 
arrows, while it remains equal to $t$ along the unmarked bonds of 
Figure III. Any reflection operation simply 
changes the signs of all the hoppings along the marked bonds. Now 
consider the unitary transformation $S$ which changes the signs of all
the Cu orbitals along both diagonal, except the central Cu. Since $S$ 
also has the effect of reversing all the arrows, $\sigma \times S$ is a 
symmetry, for all reflections $\sigma$ in $C_{4v}$.  Moreover, since 
the product of two reflections is a rotation, the Group $\tilde{C_{4v}}$
including the rotations and the reflections multiplied by $S$ is isomorphic
to $C_{4v}$. The W=0 pairs appropriate for half a flux quantum must 
involve two holes belonging to the degenerate IRREP of $\tilde{C_{4v}}$.
In this way, at  $\phi=\frac{\phi_{0}}{2}$ the full symmetry is 
restored, allowing again for  pairing and negative $\Delta$.  
The W=0 quasiparticles have just the correct symmetry properties in the 
presence of the vector potential to provide superconducting flux 
quantisation in macroscopic systems.

\section{CONCLUSIONS}

We have examined the properties of the W=0 pairs by performing
numerical diagonalizations of the $Cu_{5}O_{4}$ cluster for various 
fillings. Some of these fillings are not representative of the 
concentrations that have been realised in the cuprates, but our theory 
depends on symmetry and the concentration range to which it applies is 
wider than that obtained experimentally.  We have shown 
that the effective interaction between the two holes in the W=0 pair can be 
obtained by computing $\Delta$ by exact diagonalisation or, 
alternatively, by an analytical, recursive canonical transformation; 
we have detailed the latter approach, and derived a weak coupling approximation 
that agrees with the numerical results for $\Delta$ and with a
previous  diagrammatic analysis. Pairing occurs when $\Delta<0$.
An approximate symmetry under charge
conjugation exists leading to {\em electron pairing} as well as {\em 
hole pairing} in the sense defined in the Introduction. 
The numerical data confirm that when the filling is such that W=0 pairs are involved, 
$\Delta$ is small in absolute value, while the other fillings
lead to strong repulsion. In one case, the W=0 pair leads to a small 
repulsion, showing that the existence of pairing is not a general 
property independent on filling.  In all cases, we found that pairing 
or its absence can be reliably predicted by studying 
the behavior of the system at weak coupling, which supports the 
approximations that we performed in a study of the full plane in Ref(\cite{csb}) at last in some 
concentration ranges.  We 
stress, however, that the instability of the Fermi liquid against a 
pairing interaction does not grant superconductivity, since there is a competition with other order
parameters. Further investigations are necessary to analyse 
this point very close to half filling, where the antiferromagnetic order 
prevails and the behavior could require a strong coupling analysis. 
Moreover, we expect that the renormalisation of the 
dispersion relation cannot be neglected \cite{zanchi}.

In the $Cu_{5}O_{4}$ cluster the exact diagonalisation results show 
that W=0 pairs  quantise  flux in the superconducting way. The
ground states in presence of zero and half fluxon have different 
symmetries, like in BCS superconductors. The superconducting flux quantisation 
property is due to the fact that the symmetry Group appropriate at  
half flux quantum in isomorphic with $C_{4v}$, and this  is 
not limited to small clusters, but  general. Flux quantisation and 
pairing fit well together, both being consequences of the same 
symmetry principle.

\section{ACKNOWLEDGMENTS}
This work has been supported by the Is\-ti\-tu\-to Na\-zio\-na\-le di 
Fi\-si\-ca del\-la Ma\-te\-ria. Useful discussions with A. Sa\-gnot\-ti and R. 
Bru\-net\-ti of our  University are gratefully acknowledged.

\newpage
\begin{center}
{\bf FIGURE CAPTIONS}
\end{center}

\begin{center}
FIGURE I
\footnotesize{
 The $Cu_{5}O_{4}$ cluster with 4 flux tubes (black dots) 
	carrying flux $\phi$. 	X stands for Cu. The dotted lines represent the
	$t_{d}$ bonds providing a closed path around the centre}
\end{center} 

\bigskip

\begin{center}
FIGURE II
\footnotesize{
	Solid line (right scale):$|\left<\Psi_{0}(\phi)|\Psi_{0}(0)\right>|$. The other 
	lines (left scale) show the dimensionless response function R of 
	Equation (\ref{r})
	for 
	$n_{h}$=4,8 and 16. Note that for $n_{h}$=4, 
	$\Delta_{h}(4,\frac{\phi_{0}}{2})-\Delta_{h}(4,0) \approx 0.3 t_{d}$, 
	but $\Delta_{h}(4,\frac{\phi_{0}}{2})$ is  negative. }
\end{center} 

\bigskip
\begin{center}
FIGURE III
\footnotesize{
	Pattern of the vector potential $A$ due to 4 flux tubes (black dots) 
	carrying flux $\phi$.
	X stands for 
	Cu.The line integral of $A$ along each bond parallel to the arrow is 
	$\frac{\phi}{2}$.}
\end{center} 

\bigskip
\begin{center}
{\bf TABLES}
\end{center}
\bigskip
\begin{table}
	\begin{center}
		\begin{tabular}{lclclcl}
\hline
Symmetry &g & $\varepsilon$(eV) \\ \hline
$a_{1}$ &1 & -1.643 \\ \hline
$e$ ,$b_{1}$&3 & -0.43\\ \hline
$a_{1}$ &1 & 0. \\ \hline
$e$,$b_{1}$ &3 & 3.93 \\ \hline
$a_{1}$ &1 & 5.143 \\ \hline
\end{tabular}
\end{center}
\caption{One-hole levels of the $Cu_{5}O_{4}$ cluster, with their symmetry 
labels, degeneracies g, and energy eigenvalues $\varepsilon$  with $t=1.3$ eV, 
$\varepsilon_{p}=3.5$ eV and $\varepsilon_{d}=0$.}
\label{livelli}
\end{table}
 \smallskip
\begin{table}[tbp]
	\begin{center}

		\begin{tabular}{lclcl}
\hline
$n_{h}$ & $\Delta_{h}$ (meV) \\ \hline
4 & -15.7 \\ \hline
6 & 1469.2\\ \hline
8 & -10.85 \\ \hline
12 & 43.72 \\ \hline
14 &1109.2 \\ \hline
16 & -25.47 \\ \hline
\end{tabular}
\end{center}
\caption{Exact diagonalisation results for $\Delta_{h}(n_{h})$ (meV), using 
$U_{p}$=6 eV and $U_{d}$=5.3 eV. For $n_{h}$=4, 8 and 16 pairing takes 
place, and at $n_{h}$=12 the repulsion is drastically reduced. For $n_{h}$=6 and 14 the W=0 pairs 
are {\em not } involved and the normal repulsion develops.}
\label{meV}
\end{table}


\begin{references}
\bibitem{note} This does not imply that we are switching to a new representation 
with the vacuum at half filling; nor does it imply that the hole 
concentration or other parameters are necessarily the same as in 
hole or electron superconductors. The conventional nature of these statement is clear 
from the fact that when holes are paired, so are the electrons.
\bibitem{nota2}
In the full plane, the localised bound state is a 
superposition of many W=0 pair states, and $W_{eff}$ cannot be 
identified with  $\Delta$; one has to solve a Cooper-like equation 
to obtain the binding energy. In the small clusters like 
$Cu_{5}O_{4}$, just two W=0 configurations contribute significantly 
at weak coupling, and thus effective interaction and binding energy are the same.
\end{references}
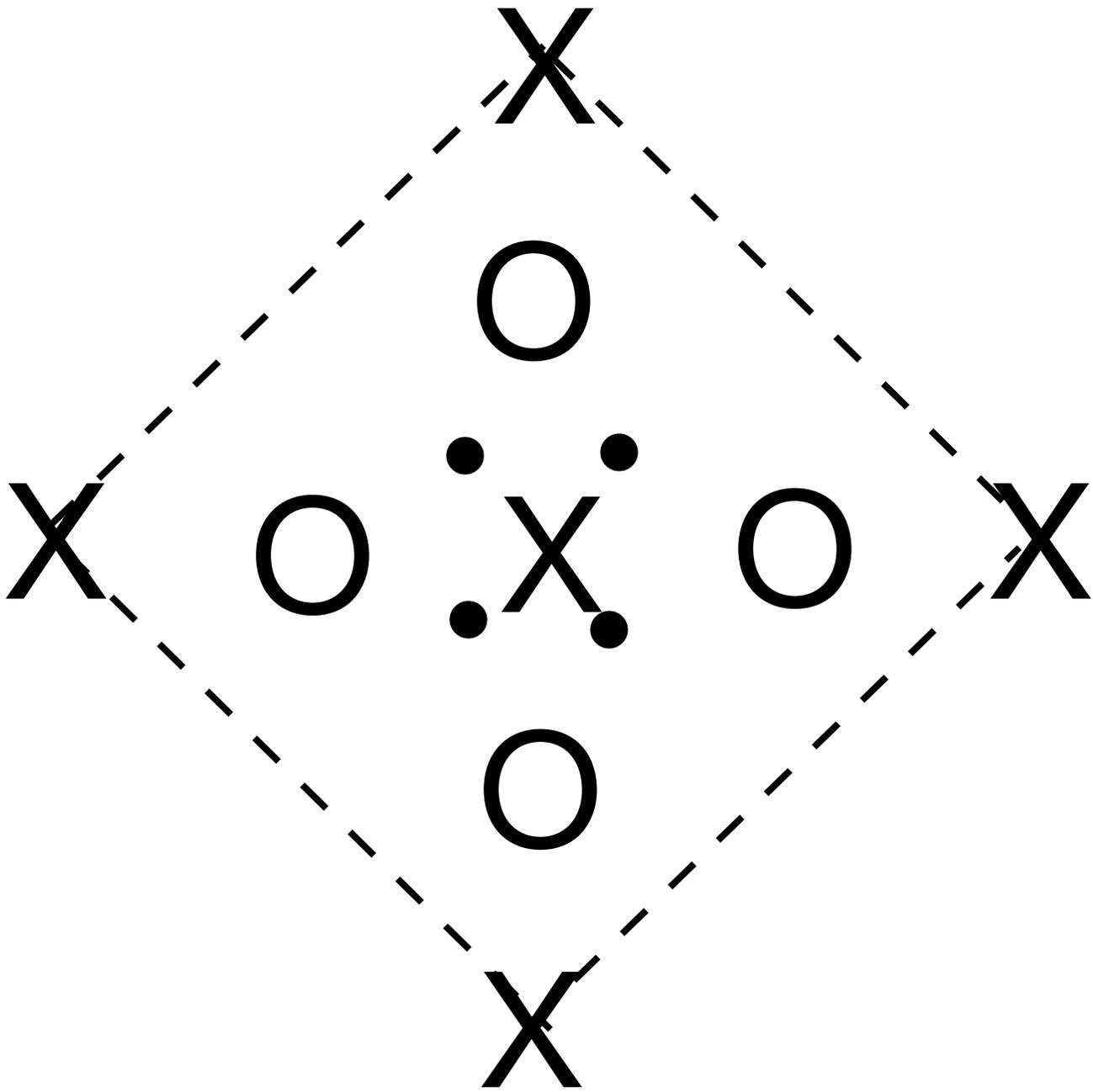
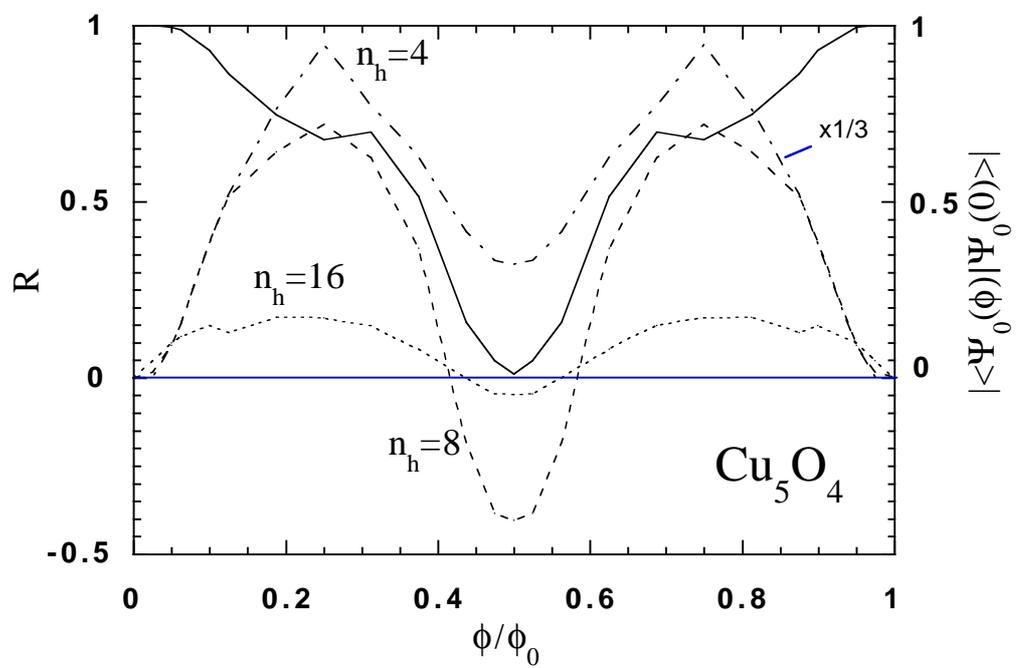
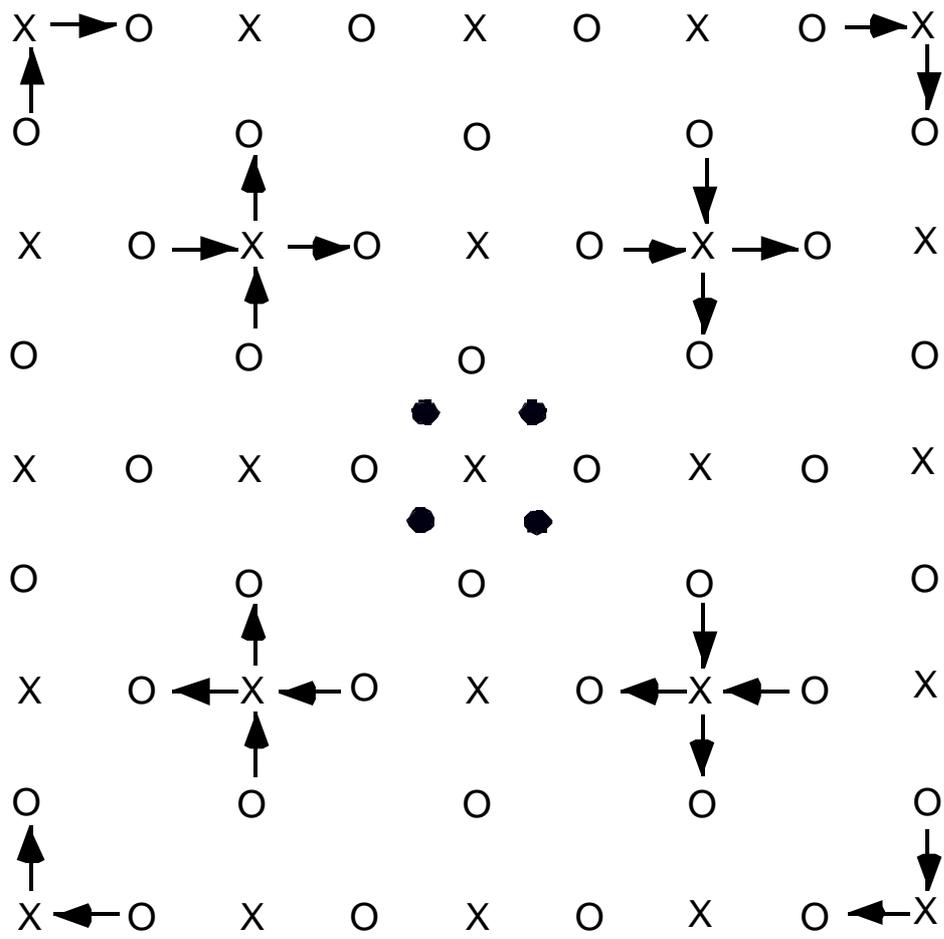
\end{document}